\documentclass{PoS}
\usepackage{amsmath}
\usepackage{url}

\newcommand{\Slash}[1]{{\ooalign{\hfil/\hfil\crcr$#1$}}}

\newcommand{\eqn}[1]{(\ref{#1})}

\def\simge{\mathrel{%
       \rlap{\raise 0.511ex \hbox{$>$}}{\lower 0.511ex \hbox{$\sim$}}}}
\def\simle{\mathrel{
       \rlap{\raise 0.511ex \hbox{$<$}}{\lower 0.511ex \hbox{$\sim$}}}}

\title{Temperature dependence of topological susceptibility using gradient flow}

\ShortTitle{Temperature dependence of topological susceptibility using gradient flow}

\author{\speaker{Yusuke Taniguchi}%
\thanks{Preprint numbers: UTCCS-P-94, UTHEP-698, KYUSHU-HET-173, J-PARC-TH-0073}
\\
Center for Computational Sciences, University of Tsukuba, Tsukuba, Ibaraki 305-8571, Japan\\
E-mail: \email{tanigchi@het.ph.tsukuba.ac.jp}}
\author{Shinji Ejiri\\
Department of Physics, Niigata University, Niigata 950-2181, Japan
}
\author{Kazuyuki Kanaya
\\
Center for Integrated Research in Fundamental Science and
Engineering (CiRfSE), University of Tsukuba, Tsukuba, Ibaraki 305-8571, Japan
}
\author{Masakiyo Kitazawa\\
Department of Physics, Osaka University, Osaka 560-0043, Japan\\
J-PARC Branch, KEK Theory Center, Institute of Particle and Nuclear Studies,
KEK, 203-1, Shirakata, Tokai, Ibaraki, 319-1106, Japan
}
\author{Hiroshi Suzuki\\
Department of Physics, Kyushu University, 744 Motooka, Fukuoka 819-0395, Japan
}
\author{Takashi Umeda\\
Graduate School of Education, Hiroshima University, Higashihiroshima, Hiroshima 739-8524, Japan
}
\author{Ryo Iwami and Naoki Wakabayashi\\
Graduate School of Science and Technology, Niigata University, Niigata 950-2181, Japan
}

\abstract{
We study temperature dependence of the topological susceptibility with the $N_{f}=2+1$ flavors
Wilson fermion.
We have two major interests in this paper.
One is a comparison of gluonic and fermionic definitions of the topological susceptibility.
Two definitions are related by the chiral Ward-Takahashi identity but their coincidence is
highly non-trivial for the Wilson fermion.
By applying the gradient flow both for the gauge and quark fields we find a good agreement of
these two measurements.
The other is a verification of a prediction of the dilute instanton gas approximation at low temperature
region $T_{pc}< T<1.5T_{pc}$, for which 
we confirm the prediction that the topological susceptibility decays with power
$\chi_{t}\propto(T/T_{pc})^{-8}$ for three flavors QCD.
}

\FullConference{34th annual ‘International Symposium on Lattice Field Theory\\
		 24-30 July 2016\\
		 University of Southampton, UK}

\begin{document}

\section{Introduction}
\label{sec:intro}

The axion is introduced into QCD through the Peccei-Quinn mechanism
to solve the strong CP problem \cite{Peccei:1977hh}.
The effective mass squared of the axion is proportional to the topological susceptibility
and its temperature dependence plays a crucial role for the axion production
in the early universe at high temperature and for a possibility of the axion to be
a candidate of cold dark matter.

Recently the topological susceptibility is studied at finite temperature using lattice QCD 
for quenched case \cite{Berkowitz:2015aua,Borsanyi:2015cka,Kitano:2015fla},
for $N_{f}=2+1$ flavours \cite{Bonati:2015vqz,Petreczky:2016vrs} and
for $N_{f}=2+1+1$ flavors \cite{Borsanyi:2016ksw}.
One of major interests of these papers is consistency with the dilute instanton gas approximation
(DIGA) \cite{Gross:1980br}, which predicts a power law decay of the topological susceptibility
$\chi_{t}\propto(T/T_{pc})^{-8}$ at high temperature for three flavors.
The result of Ref.~\cite{Bonati:2015vqz} is the decay is much more gentle than DIGA prediction.
On the other hand Ref.~\cite{Petreczky:2016vrs} shows the power is consistent with that of DIGA
above $1.5T_{pc}$ but it is a bit more moderate for low temperature region $T_{pc}< T<1.5T_{pc}$.
In this paper we focus on temperature dependence of the topological susceptibility at
$T_{pc}< T<1.5T_{pc}$ for $N_{f}=2+1$ QCD.
Our result is the power is consistent with that of DIGA even at low temperature region.

One of the best way to measure the topological susceptibility may be to use lattice QCD.
However there are several definitions for the topological charge on lattice.
The most popular one may be to adopt  the gauge field strength $F\tilde{F}$
accompanied with a cooling step
\cite{GarciaPerez:1998ru,Bonati:2014tqa,Namekawa:2015wua,Alexandrou:2015yba}.
In this paper we use the definition by adopting the gradient flow 
\cite{Luscher:2009eq,Luscher:2010iy,Narayanan:2006rf}.
The gauge field is flowed with fictitious time $t$ according to the equations \cite{Luscher:2010iy}
\begin{equation}
   \partial_tB_\mu(t,x)=D_\nu G_{\nu\mu}(t,x),\quad
   B_\mu(t=0,x)=A_\mu(x),
\label{eq:(1.1)}
\end{equation}
where the field strength $G_{\nu\mu}(t,x)$ and the covariant derivative $D_{\nu}$ %in the equation
are given in terms of the flowed gauge field.
%\begin{eqnarray}
%G_{\mu\nu}(t,x)&=&
%\partial_\mu B_\nu(t,x)-\partial_\nu B_\mu(t,x)+[B_\mu(t,x),B_\nu(t,x)],
%\label{eq:(1.2)}
%\\
%D_\nu G_{\nu\mu}(t,x)&=&
%\partial_\nu G_{\nu\mu}(t,x)+[B_\nu(t,x),G_{\nu\mu}(t,x)].
%\end{eqnarray}
%The gauge field is smeared within a range of $\sqrt{8t}$ and the flow works as a kind of cooling.
The topological charge density~$q(t,x)$ defined by the flowed gauge field
\begin{equation}
   q(t,x)=\frac{1}{64\pi^2}
   \epsilon_{\mu\nu\rho\sigma}G_{\mu\nu}^a(t,x)G_{\rho\sigma}^a(t,x),
\quad
\epsilon_{0123}=1
\label{eq:(1.4)}
\end{equation}
is already renormalized \cite{Luscher:2010iy} and its normalization is consistent with
the Ward-Takahashi  (WT)  identity associated with the flavor singlet chiral symmetry
\cite{Ce:2015qha}.
The topological susceptibility is given by
\begin{eqnarray}
 \chi_{t}=\frac{1}{V_4}\left(\left\langle Q^2 \right\rangle
-\left\langle Q\right\rangle^2\right)
=\frac{1}{V_4}\left\langle Q^2 \right\rangle,
\quad
Q(t)=\int d^4x\,q(t,x).
\label{eq:(1.5)}
\end{eqnarray}

In the continuum QCD the topological susceptibility is related to the disconnected
singlet pseudo-scalar two point function
\cite{Bochicchio:1984hi,Giusti:2004qd}
through the chiral WT identity
\begin{eqnarray}
\left\langle \partial_\mu A_\mu^a(x) {\cal O}\right\rangle
-2m\left\langle \pi^a(x) {\cal O}\right\rangle
+2n_f\delta^{a0}\left\langle q(x) {\cal O}\right\rangle
=i\left\langle \delta^a{\cal O}\right\rangle,
&&
\end{eqnarray}
where $a=0$ stands for the singlet and $a\ge1$ for the non-singlet identity.
$A_{\mu}^a(x)=\bar\psi(x)T^a\gamma_{\mu}\gamma_5\psi(x)$,
$\pi^a(x)=\bar\psi(x)T^a\gamma_5\psi(x)$ with $T^0=1$ and
${\rm tr}\left(T^aT^b\right)=\delta^{ab}$ for $a,b\ge1$.
$n_{f}$ is a number of flavors with degenerate mass $m$, which is not same as number of sea quarks
necessarily.
We briefly explain the derivation in the following.
% in order to be self-contained.
We apply the integrated form of the singlet WT identify to ${\cal O}=Q$ and ${\cal O}=P^{0}$
\begin{eqnarray}
&&
-m\left\langle P^0 Q\right\rangle
+n_f\left\langle Q^2 \right\rangle
=0,
\\&&
-m\left\langle P^0 P^0\right\rangle
+n_f\left\langle Q P^0\right\rangle
=-\left\langle S^0\right\rangle
\end{eqnarray}
and the non-singlet WT identity to non-singlet ${\cal O}=P^{b}$
\begin{eqnarray}
-2m\left\langle P^a P^b\right\rangle
%&=&
%i\left\langle \delta^aP^b\right\rangle
=-\left(\delta^{ab}\frac{2}{n_f}\left\langle S^0\right\rangle
+d_{abc}\left\langle S^c\right\rangle
\right),
\quad
a,b,c\ge1,
\end{eqnarray}
where
$P^a=\int d^4x \pi^a(x)$ and $S^a=\int d^4x \bar\psi(x)T^a\psi(x)$.
%And we have
%\begin{eqnarray}
%&&
%n_f^2\left\langle Q^2 \right\rangle
%=m^2\left\langle P^0 P^0\right\rangle-m\left\langle S^0\right\rangle
%\end{eqnarray}
Making use of a fact that the non-singlet flavor symmetry is not broken we get
\begin{eqnarray}
&&
\left\langle Q^2 \right\rangle
=\frac{m^2}{n_f^2}\left(
\left\langle P^0 P^0\right\rangle-n_f\left\langle P^a P^a\right\rangle
\right),
%=\frac{m^2}{n_f^2}
%\left\langle P^0 P^0\right\rangle_{\rm disconnecterd},
\label{eqn:disconnected}
\end{eqnarray}
where sum is not taken over $a$.
The right hand side is nothing but the disconnected contribution to the
singlet pseudo-scalar two point function.
%The expectation value of the topological charge is given by
%\begin{eqnarray}
%\left\langle Q\right\rangle=\frac{m}{n_{f}}\left\langle P^{0}\right\rangle.
%\end{eqnarray}

The right hand side of \eqn{eqn:disconnected} would have power divergence when calculated
on lattice using the Wilson fermion since the chiral symmetry is broken explicitly.
Much efforts were payed to avoid the difficulty \cite{Luscher:2004fu,Giusti:2008vb,Luscher:2010ik}.
In this paper we shall use a new method to get rid of the singularity.
This is accomplished by applying the gradient flow to the quark fields \cite{Luscher:2013cpa}
\begin{eqnarray}
&&
\partial_t\chi_f(t,x)=\Delta\chi_f(t,x),
\quad
   \chi_f(t=0,x)=\psi_f(x),
\label{eq:(1.14)}
\\&&
\partial_t\Bar{\chi}_f(t,x)
   =\Bar{\chi}_f(t,x)\overleftarrow{\Delta},
   \quad
   \Bar{\chi}_f(t=0,x)=\Bar{\psi}_f(x)
\label{eq:(1.15)}
\end{eqnarray}
with
\begin{eqnarray}
&&
\Delta\chi_f(t,x)\equiv D_\mu D_\mu\chi_f(t,x),
\label{eq:(1.16)}
\quad
D_\mu\chi_f(t,x)\equiv\left[\partial_\mu+B_\mu(t,x)\right]\chi_f(t,x),
\\&&
\Bar{\chi}_f(t,x)\overleftarrow{\Delta}
   \equiv\Bar{\chi}_f(t,x)\overleftarrow{D}_\mu\overleftarrow{D}_\mu,
\label{eq:(1.17)}
\quad
   \Bar{\chi}_f(t,x)\overleftarrow{D}_\mu
   \equiv\Bar{\chi}_f(t,x)\left[\overleftarrow{\partial}_\mu-B_\mu(t,x)\right],
\end{eqnarray}
where $f=u$, $d$, $s$, denotes the flavor index.
It is probed that any operator constructed with the flowed quark field does not have any UV divergence
when multiplied with a wave function renormalization factor of the quark field \cite{Luscher:2013cpa}.
%The flowed operator $\overline{\chi}(t,x)\gamma_{5}\chi(t,x)$ is renormalized non-perturbatively
We adopt the wave function renormalization factor given by Ref.~\cite{Makino:2014taa}
\begin{equation}
   \varphi_f(t)\equiv
   \frac{-6}
   {(4\pi)^2t^2
   \left\langle\Bar{\chi}_f(t,x)\overleftrightarrow{\Slash{D}}\chi_f(t,x)
   \right\rangle_{\! 0}},
\quad
   \overleftrightarrow{D}_\mu\equiv D_\mu-\overleftarrow{D}_\mu,
\label{eq:(1.18)}
\end{equation}
where expectation value is taken at zero temperature.
In the end we need to convert the renormalized operator
$\varphi_{f}(t)\overline{\chi}_{f}(t,x)\gamma_{5}\chi_{f}(t,x)$
to the pseudo-scalar density which is consistent with the chiral WT identity.
This is accomplished perturbatively according to the strategy of Ref.~\cite{Hieda:2016lly}
based on a small flow time expansion \cite{Luscher:2011bx}
\begin{eqnarray}
m_{R}\left(\overline{\psi}_{f}\gamma_{5}\psi_{f}\right)_{R}=\lim_{t\to0}
&&
c_{S}(t)\Bar{m}_{\overline{\rm MS}}(1/\sqrt{8t})
\varphi_{f}(t)\overline{\chi}_{f}(t,x)\gamma_{5}\chi_{f}(t,x),
\label{eqn:cpp}
\end{eqnarray}
where
\begin{equation}
c_{S}(t)=1+\frac{\Bar{g}_{\overline{\rm MS}}(1/\sqrt{8t})^2}{(4\pi)^2}
   \left[4\,(\gamma-2\ln2)
   +8
   +\frac{4}{3}\,\ln(432)\right]
\end{equation}
is the matching coefficient evaluated in ${\overline{\rm MS}}$ scheme.
$\Bar{g}_{\overline{\rm MS}}$ and $\Bar{m}_{\overline{\rm MS}}$ are the running coupling and mass
in ${\overline{\rm MS}}$ scheme at renormalization scale $\mu=1/\sqrt{8t}$.
Notice that the scheme dependence is canceled out and the left hand side of \eqn{eqn:cpp} is scheme and scale independent.
%Our strategy works well and the gauge and fermion definition is
%consistent with each other at temperature $T/T_{pc}<1.5$.

%%%%%%%%%%%%%%%%%%%%%%%%%%%%%%%%%%%%%%%%%%%%%%%%%%%%
\section{Simulation parameters}
\label{sec:parameters}

Measurements are performed on $N_f=2+1$ gauge
configurations generated for Refs.~\cite{Ishikawa:2007nn,Umeda:2012er},
which are open to the public on ILDG/JLDG. %~\cite{Maynard:2010wi}.
A non-perturbatively ${O}(a)$ improved Wilson quark
action and the renormalization-group
improved Iwasaki gauge action are adopted. The bare coupling
constant is set to $\beta=2.05$, which corresponds
to~$a=0.0701(29)\,\mathrm{fm}$ ($1/a\simeq2.79\,\mathrm{GeV}$).
%The non-perturbative clover coefficient is $c_{\mathrm{SW}}=1.628$ at~$\beta=2.05$.
The hopping parameters are set to $\kappa_u=\kappa_d\equiv\kappa_{ud}=0.1356$ and $\kappa_s=0.1351$,
which correspond to heavy up and down quarks, $m_\pi/m_\rho\simeq0.63$, and
almost physical strange quark, $m_{\eta_{ss}}/m_\phi\simeq0.74$. The bare PCAC
quark masses are
$a\,m_{ud}=0.02105(17)$ and $a\,m_s=0.03524(26)$.

In this study, we adopt the fixed-scale approach~\cite{%Levkova:2006gn,%
Umeda:2008bd} in which the temperature $T=1/(aN_t)$ is varied by changing the
temporal lattice size $N_{t}$ with a fixed lattice spacing~$a$.
% so that a common parameter set of zero-temperature results can be used for all temperatures.
We adopt $4\le N_t\le16$ which correspond to $174\simle T\simle697$ (MeV).
See Ref.~\cite{Taniguchi:2016ofw} for temperature and number of configurations at each $N_{t}$.
Spatial box size is $32^3$ for finite temperature and $28^3$ for zero temperature.
%The physical box size is all the same for~$T>0$.

To evaluate fermionic observables we use the noisy estimator method. 
The number of noise vectors is 20 for each color.
%To compute observables at $t>0$, we need flowed gauge and quark fields. 
We adopt the third order Runge-Kutta method~\cite{Luscher:2010iy,Luscher:2013cpa} with the step
size of~$\epsilon=0.02$ to solve the flow equation for both the gauge and quark fields. 

%The statistical errors are estimated by the standard jackknife analysis.
For a study of the auto-correlation we perform the bin size analysis of
the jackknife error.
We find that bin size of $300$ in Monte Carlo time is enough for the statistical
error to saturate.
For the quadratic terms of the field strength tensor $G_{\mu\nu}(x)$
we adopt a combination of the clover operator with four plaquette Wilson loops 
and that with four $1\times2$ rectangle Wilson loops such that 
the tree-level improved field strength squared is
obtained~\cite{AliKhan:2001ym}.
The fermionic definition is applied to $n_{f}=2$ flavors $ud$ quark sub-system,
which has degenerate mass.

%%%%%%%%%%%%%%%%%%%%%%%%%%%%%%%%%%%%%%%%%%%%%%%%%%%%
\section{Numerical results}
\label{sec:results}

%\subsection{Gluonic definition}

\begin{figure}[h]
 \begin{center}
  \includegraphics[width=4.7cm]{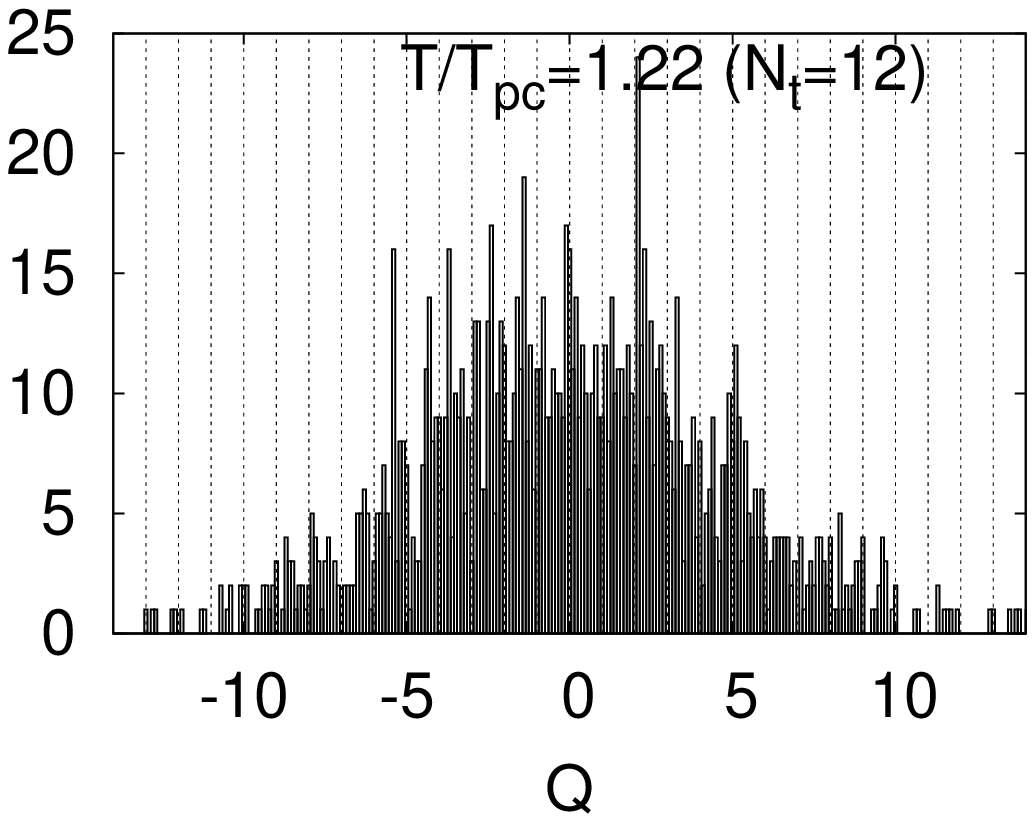}
  \includegraphics[width=4.7cm]{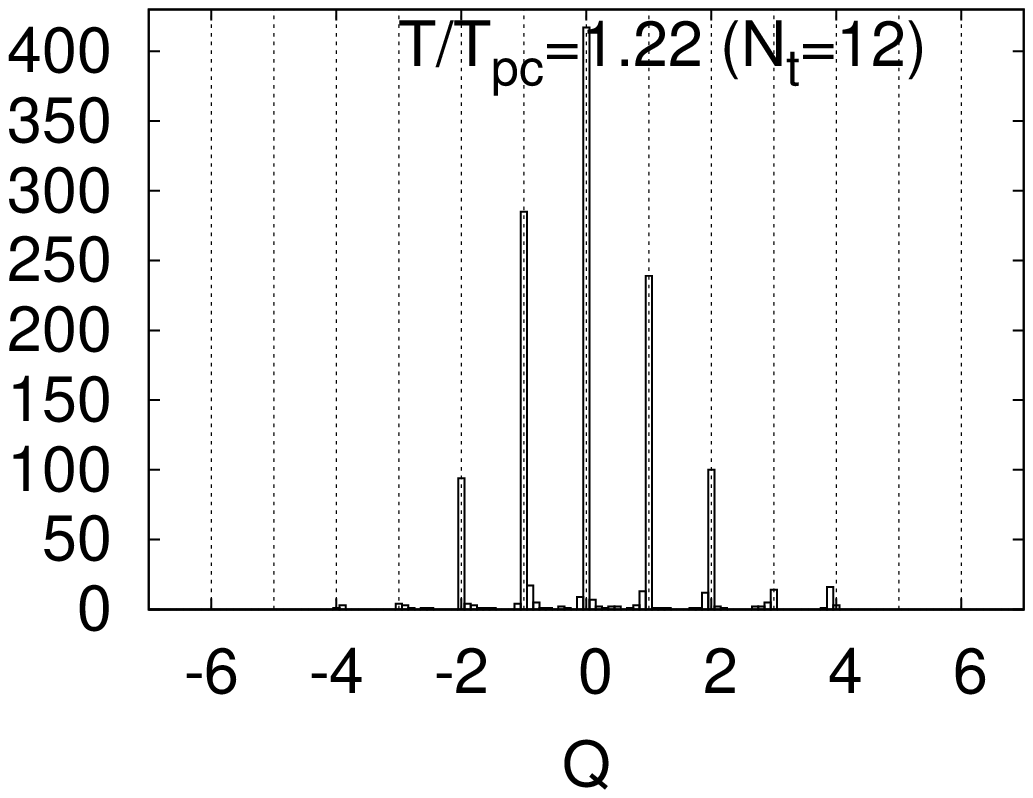}
  \includegraphics[width=4.7cm]{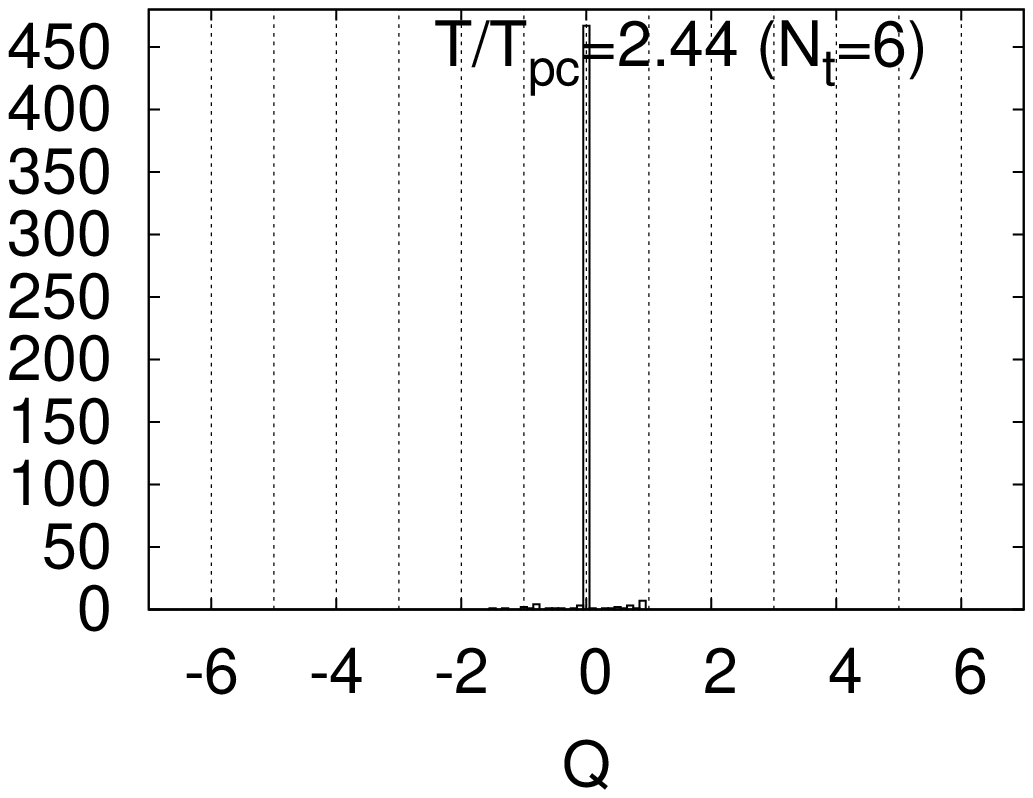}
  \vspace*{-1em}
  \caption{Distribution of the topological charge for $T/T_{pc}\simeq1.22$ at flow time $t/a^{2}=0.02$
  (left panel), $4.5$ (middle panel) and for $T/T_{pc}\simeq2.44$ at $t/a^{2}=1.125$ (right panel).
  Dotted vertical lines indicate integers.
}
\label{eqn:histogram}
 \end{center}
\end{figure}

In Fig.~\ref{eqn:histogram} we plot a distribution of the topological charge by the gluonic definition
\eqn{eq:(1.4)} for $T/T_{pc}\simeq1.22$ at flow time $t/a^{2}=0.02$ (left panel) and 
$4.5$ (middle).
We can see dense and wide distribution in the left panel
are accumulated on integer values in the middle panel as we flow the gauge field with a large enough time.
We stop the flow before we reach the over-smeared region
$t_{1/2} \equiv \frac{1}{8}\left[\min\left(\frac{N_t}{2},\frac{N_s}{2}\right)\right]^2$,
where the typical smearing range $\sqrt{8t}$ of the gradient flow covers smaller side of the
lattice box.
The topological charges are well distributed on non-zero values for $0\simle T/T_{pc}\simle1.47$ but the fluctuation becomes rare for high temperature region $T/T_{pc}\simge2.44$ as is shown in the right panel.

\begin{figure}[h]
 \begin{center}
  \includegraphics[width=5.5cm]{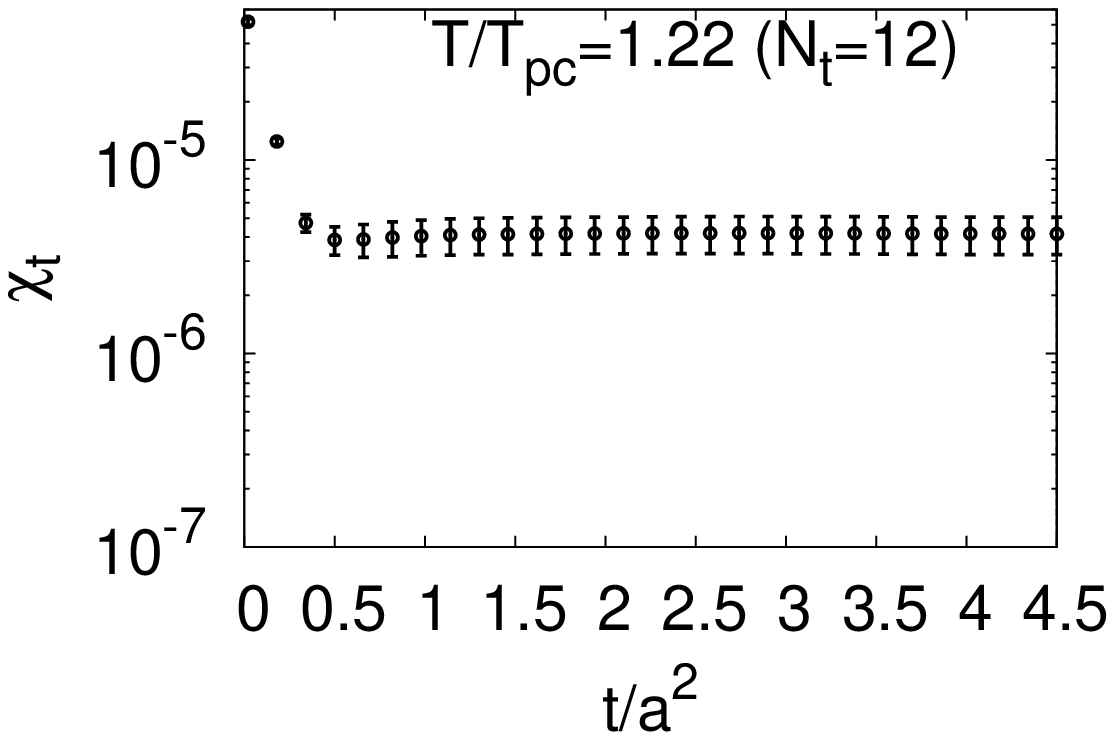}
  \includegraphics[width=5.5cm]{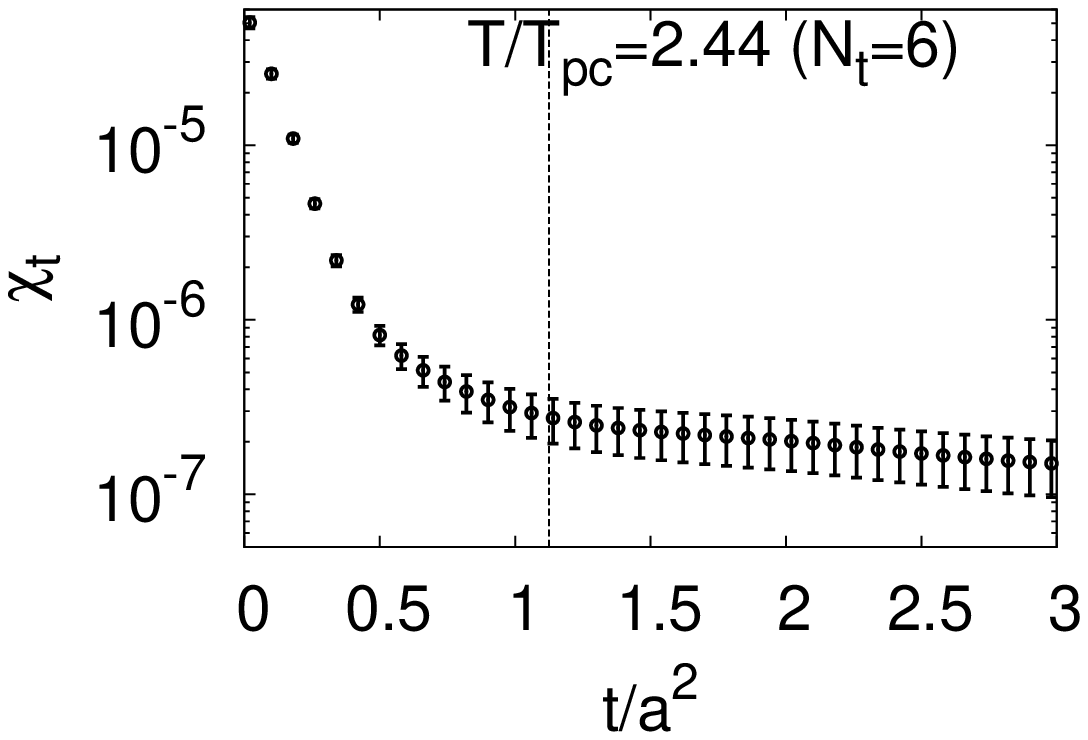}
  \vspace*{-2ex}
  \caption{Topological susceptibility as a function of the flow time $t/a^2$ for
   $T/T_{pc}\simeq1.22$ (left panel) and $T/T_{pc}\simeq2.44$ (right panel).
   Dotted vertical line in the right panel indicates $t_{1/2}$.
}
\label{eqn:Q_susceptibility_vs_t}
 \end{center}
\end{figure}

Fig.~\ref{eqn:Q_susceptibility_vs_t} is the topological susceptibility as a function of the flow time for
$T/T_{pc}\simeq1.22$ (left panel) and $T/T_{pc}\simeq2.44$ (right), which is defined by the gluonic operator
\eqn{eq:(1.4)}.
The susceptibility is completely flat for $T/T_{pc}\simeq1.22$ (left panel) at large flow time.
This is consistent with the flow time invariant property of the topological charge in the continuum limit
\cite{Ce:2015qha}.
%This is consistent with a fact that the gradient flow plays a role of renormalization whose scale is
%given by $\mu=1/\sqrt{8t}$ and the topological charge is scale invariant.
The good property is observed for $T/T_{pc}\simle1.47$ and we adopt the value at $t_{1/2}$
as our result.
On the other hand the topological susceptibility does not have a plateau even above $t_{1/2}$
for $T/T_{pc}\simge1.83$ as is shown in the right panel for $T/T_{pc}\simeq2.44$.
This is supposed to be mainly due to the lattice artifact $aT=1/N_{t}$, which becomes severe at
high temperature.
We give up to study the high temperature region $T/T_{pc}\simge1.83$ with the gluonic definition
in this paper.

%\subsection{Fermionic definition}

We consider the fermionic definition of the topological susceptibility
using \eqn{eqn:disconnected} with the gradient flow renormalization
\eqn{eqn:cpp}.
In this procedure we need to take two limits in a proper order;
the continuum limit first and $t\to0$ limit later.
The continuum limit is usually taken by an extrapolation from more than three lattice spacings
around $a\sim1$ fm.
It should be argued whether enough information is given at such lattice spacings for taking $t\to0$ limit.
For this purpose we consider the lattice spacing and the flow time dependence of
the topological susceptibility $\chi_{t}$
\begin{eqnarray}
\chi_{t}(t,a)&=&
\chi_{t}+A\frac{a^2}{t}+tS
+\sum_{f}B_f(am_f)^2+C(aT)^2
%\nn\\&&
+D\left(a\Lambda_{\rm QCD}\right)^2
+a^2S'+{O}(a^4,t^2),
\label{eqn:expansion}
\end{eqnarray}
where 
$A$, $B$, $C$, $D$ are contributions from four dimensional operators and
$S$, $S'$ are those from dimension six operators.
The second term $a^{2}/t$ is the reason why we need to keep the proper order of the limit.
However this term becomes negligibly small at large $t/a^2$.
On the other hand ${O}(t^2)$ term becomes dominant at such a flow time.
Our conclusion is that we can exchange the order of the limit and take $t\to0$ limit first
if there is a window region where both effects are negligible and
the data behaves linearly in $t/a^2$ .
The term $tS$ is the reason why we need to take $t\to0$ limit.

\begin{figure}[h]
 \begin{center}
  \includegraphics[width=4.9cm]{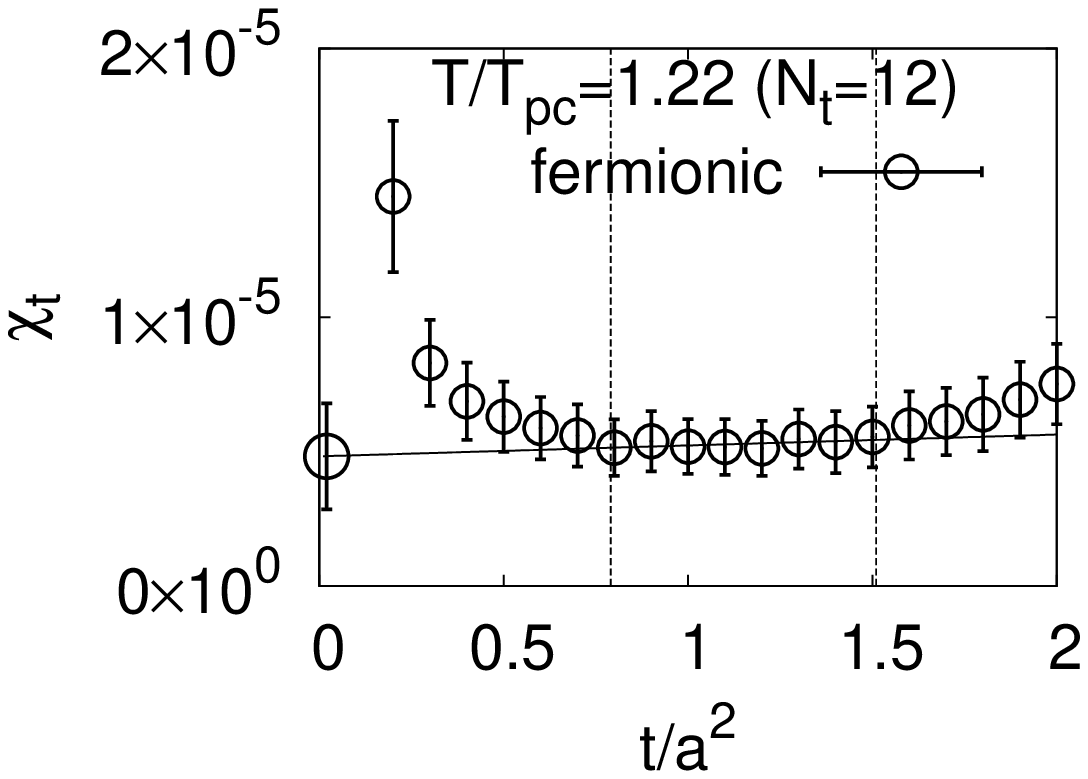}
  \hspace*{-5mm}
  \includegraphics[width=4.9cm]{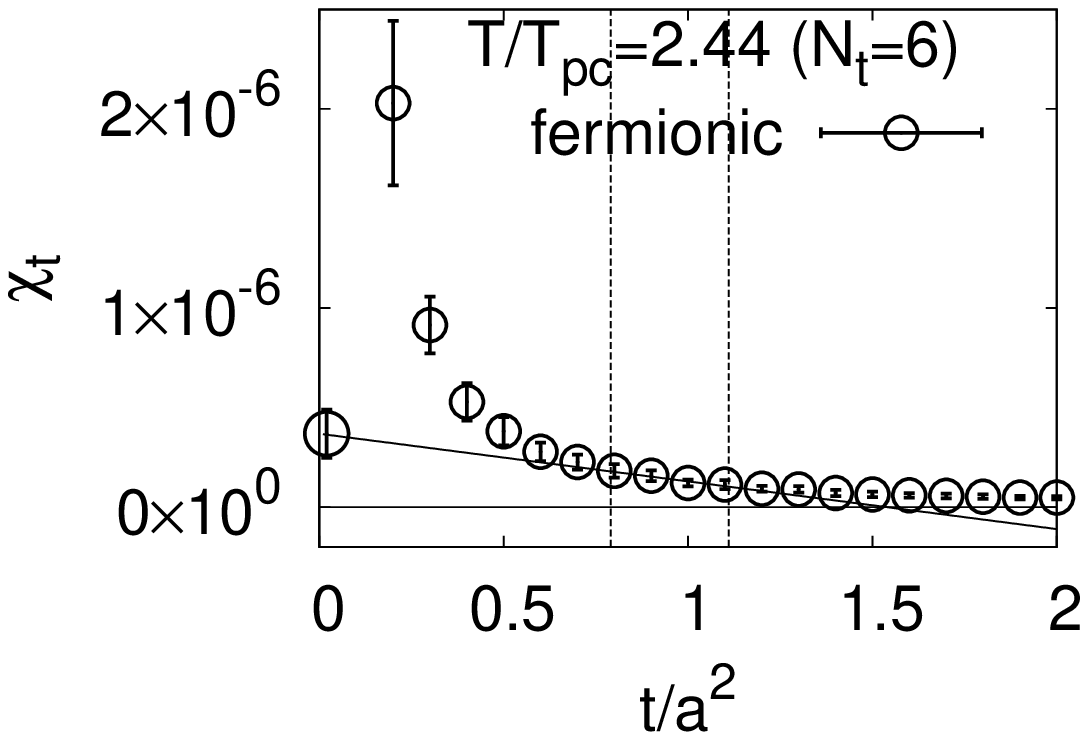}
  \hspace*{-5mm}
  \includegraphics[width=5.8cm]{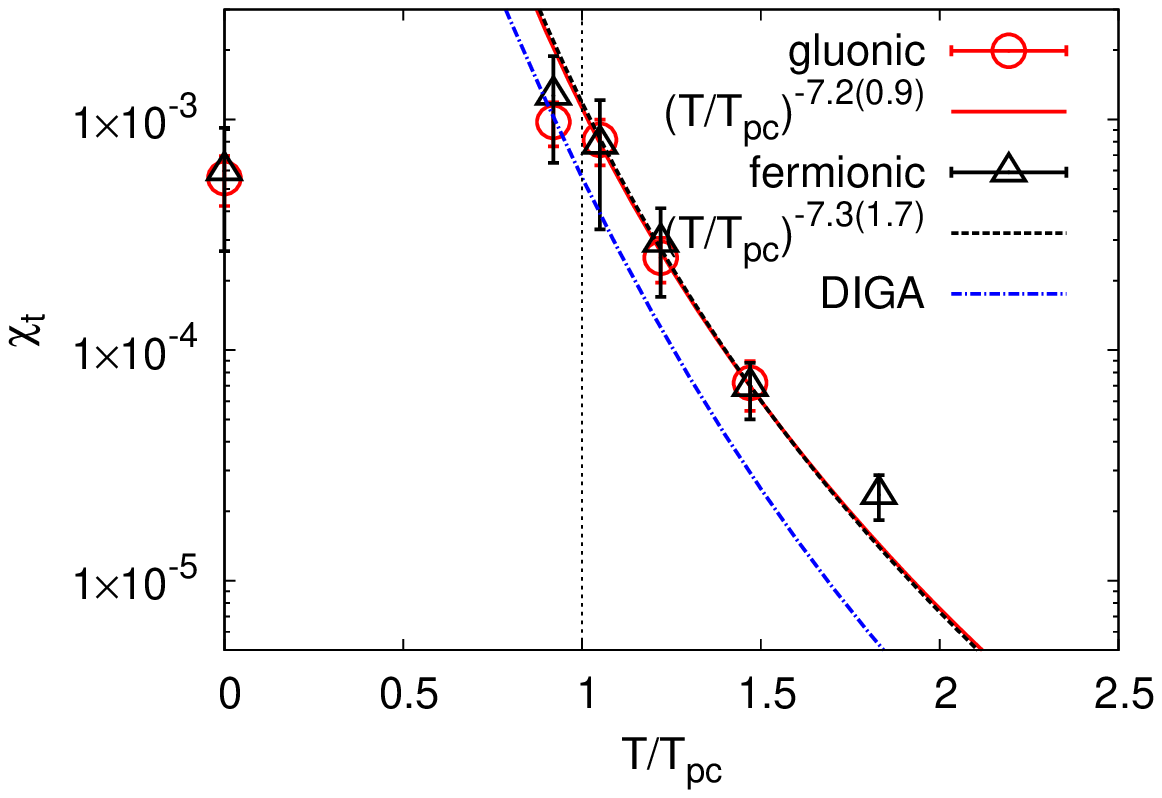}
  \vspace*{-2ex}
  \caption{Topological susceptibility as a function of the flow time
  $t/a^2$ for $T/T_{pc}\simeq1.22$ (left panel), $2.44$ (middle panel).
The right panel is the topological susceptibility in a unit of $({\rm GeV})^4$ as a function of temperature,
where gluonic and fermionic definitions are compared with DIGA result.
Red and black lines are fit of gluonic and fermionic results above $T_{pc}$.
Dotted blue line is a prediction of DIGA.
}
\label{fig:Q_susceptibility_gauge_vs_fermion}
 \end{center}
\end{figure}
In Fig.~\ref{fig:Q_susceptibility_gauge_vs_fermion} we plot $\chi_{t}(t,a)$ as a function of the flow
time for $T/T_{pc}\simeq1.22$ (left panel) and $2.44$ (middle).
The non-linear behavior near the origin may be due to the lattice
artifact $a^2/t$ and that at large flow time may be ${O}(t^2)$
contribution.
We find a rigid window for $T/T_{pc}\simle1.83$ indicated by the vertical lines,
which is set to be common for a calculation of the chiral condensate and susceptibility
in Ref.~\cite{Taniguchi:2016ofw}.
The $t\to0$ limit is taken by a linear fit in the window.
We notice the window should be well below the over smeared region $t_{1/2}$.
Unfortunately the window is obscure for $T/T_{pc}\simge2.44$ mainly due to small $N_{t}$
and we could not get valid result at high temperature.

The right panel of Fig.~\ref{fig:Q_susceptibility_gauge_vs_fermion} is our result, where
the gluonic and fermionic definition of the topological susceptibility are plotted as a function
of temperature.
The results from both the definitions are consistent with each other
below $T/T_{pc}\simle1.47$.
%The deviation at high temperature $T/T_{pc}\simge2.44$ may be due to a lattice
%artifact of the form $aT=1/N_t$, which is given by $C(aT)^2$ in
%\eqn{eqn:expansion} and may appear as no plateau property (explanation
%postponed) for gluonic definition.
We fit the data at $T/T_{pc}\simeq1.05, 1.22, 1.47$ with a power of $(T/T_{pc})^\gamma$.
We have $\gamma=-7.2(0.9)$ for the gluonic and $\gamma=-7.3(1.7)$ for
the fermionic definition.
These exponents are consistent with the prediction $\gamma=8$ of DIGA in the high temperature limit
within statistical error.
Result of DIGA is also plotted by dotted blue line, where we adopted the same bare quark mass
and $T_{pc}\sim190$ MeV as our simulation for the input.
Although the exponent is consistent, our numerical result is $1.8$ times larger at
$T/T_{pc}\simeq1.22$.

%%%%%%%%%%%%%%%%%%%%%%%%%%%%%%%%%%%%%%%%%%%%%%%%%%%%
\section{Conclusions and discussions}
\label{sec:conclusion}

We study temperature dependence of the topological susceptibility from
two interests.
One is to compare two independent measurements of the susceptibility on
lattice with Wilson fermion.
We calculate the topological susceptibility adopting the gluonic
\eqn{eq:(1.4)} and fermionic \eqn{eqn:disconnected} definitions,
for which we apply the gradient flow.
Although the gradient flow is used as a renormalization for both definitions
the procedure to extract the topological susceptibility is different.
% according to the different strategies.
%The gradient flow is used as a cooling for the gluonic definition and
%plays a role of renormalization for the fermionic definition.
The independent results for two definitions agree perfectly well for $T/T_{pc}\simle1.47$.
%Discrepancy at high temperature $T/T_{pc}\simge2.44$ may be due to a lattice
%artifact.

The other is a test of the dilute instanton gas approximation prediction
at low temperature region $T_{pc}\simle T\simle1.5T_{pc}$.
%that the topological susceptibility decays with power
%$\chi_{t}\propto(T/T_{pc})^{-8}$ at high temperature.
By fitting the lowest three data above $T_{pc}$ with a power law
$\chi_{t}\propto(T/T_{pc})^{\gamma}$ the exponent is consistent with the DIAG
prediction for both the definitions.
The absolute value is about two times larger than the DIGA result.
In this paper we adopt a rather heavy $ud$ quark mass with
$m_\pi/m_\rho\simeq0.63$.
In our future work we shall make use of $ud$ quark mass at the physical
point and shall discuss the axion abundance in a realistic manner.

\acknowledgments
This work is in part supported by JSPS KAKENHI Grant
No.\ 25800148, No.\ 26287040, No.\ 26400244, No.\ 26400251, No.\ 15K05041,
and No.\ 16H03982,
by the Large Scale Simulation Program of High Energy Accelerator
Research Organization (KEK) No.\ 14/15-23, 15/16-T06, 15/16-T-07, and 15/16-25, 
and by Interdisciplinary Computational Science Program in CCS, University of
Tsukuba.
This work is in part based on Lattice QCD common code Bridge++ \cite{bridge}.

%%%%%%%%%%%%%%%%%%%%%%%%%%%%%%%%%%%%%%%%%%%%%%%%%%%%

\end{document}